\documentclass[12pt,onecolumn]{IEEEtran}
\usepackage[colorlinks,bookmarksopen,bookmarksnumbered,citecolor=red,urlcolor=red,]{hyperref}
 \usepackage{indentfirst}
\usepackage{algorithm,algorithmic,amsbsy,amsmath,amssymb,epsfig,bbm,mathrsfs, bbm} 
\usepackage{multirow}
\usepackage{mathrsfs}
 \usepackage{epstopdf}

\usepackage{verbatim}
\usepackage{amsfonts}
\usepackage{amsthm}
\begin{document}

\title{Resource Allocation in Wireless Networks with RF Energy Harvesting and Transfer}
\author{Xiao Lu$^1$, Ping Wang$^1$, Dusit Niyato$^1$, and Zhu Han$^2$	\\
$^1$ School of Computer Engineering, Nanyang Technological University (NTU), Singapore\\
$^2$ Electrical and Computer Engineering, University of Houston, Texas, USA. \vspace{-5mm}
\thanks{\textbf{D. Niyato} is the corresponding author.}}


\markboth{\emph{IEEE Network}}
{Shell \MakeLowercase{\textit{et al.}}: Bare Demo of IEEEtran.cls for Journals}

\maketitle

\begin{abstract}
Radio frequency (RF) energy harvesting and transfer techniques have recently become alternative methods to power the next generation of wireless networks. As this emerging technology enables proactive replenishment of wireless devices, it is advantageous in supporting applications with quality-of-service (QoS) requirement. This article focuses on the resource allocation issues in wireless networks with RF energy harvesting capability, referred to as RF energy harvesting networks (RF-EHNs). First, we present an overview of the RF-EHNs, followed by a review of a variety of issues regarding resource allocation. Then, we present a case study of designing in the receiver operation policy, which is of paramount importance in the RF-EHNs. We focus on QoS support and service differentiation, which have not been addressed by previous literatures. Furthermore, we outline some open research directions. 
\end{abstract}

\section{Introduction}
 
Recently, there has been an upsurge of research interests in radio frequency (RF) energy harvesting/scavenging technique (see~\cite{Visser2013} and references therein), which is the capability of converting the received RF signals into electricity. This technique has become a promising solution to power energy-constrained wireless networks. Conventionally, the energy-constrained wireless networks, such as wireless sensor networks, have a limited lifetime, which significantly confines the network performance. In contrast, an RF energy harvesting network (RF-EHN) can have power supply from a radio environment. Consequently, RF-EHNs have found their applications quickly in various forms, such as wireless sensor networks~\cite{Nishimoto2010}, wireless body networks~\cite{Zhang2010}, and wireless charging systems. For example, \cite{Nishimoto2010} presents a prototype implementation of sensor nodes powered by ambient RF energy. In~\cite{Zhang2010}, the authors design an RF-powered integrated circuit with a work-on-demand protocol for wireless body networks in medical applications. With the increasingly emerging applications of RF energy harvesting/charging, the Wireless Power Consortium (www.wirelesspowerconsortium.com) is also making efforts of establishing an international standard for the RF energy harvesting and transfer technology. Note that the RF energy harvesting typically refers to the capability of the wireless devices to harvest energy from RF signals. The RF energy transfer refers to as the method and mechanism of an RF source to transmit RF energy to the wireless devices. 
 
In RF energy harvesting, radio signals with frequency range from $300$ GHz to as low as $3$ KHz are used as a medium to carry energy in a form of electromagnetic radiation. Wireless information is modulated on the amplitude and phase of RF waves, while wireless energy transfer is carried out through far-field RF radiation. RF energy transfer is characterized by low-power and long-distance transfer, and thus is suitable for powering a large number of devices with low energy consumption, dispersed in a relatively wide area. Due to the specific nature of the RF energy harvesting and wireless communication requirements, wireless networks have to be re-designed to achieve maximal efficiency for RF energy harvesting and transfer. In particular, the resource allocation for wireless networks has to be optimized considering the tradeoff among network performance, energy efficiency, and RF energy supply.

This article presents recent advances in RF-EHNs. We first provide an overview of the RF-EHNs. Then, we introduce and discuss about different resource allocation issues. Furthermore, we highlight the importance of a receiver operation policy by showing a case study in a general RF-EHN. Realizing that none of previous works in the literature considers the receiver operation problem with service differentiation, we aim to fill the gap. Specifically, we devise an optimal operation policy that provides service differentiation among different types of traffic, i.e., low priority (LP) and high priority (HP) data, as well as to meet their quality-of-service (QoS) requirements. We formulate an optimization model and obtain the optimal operation policy that maximizes the weighted sum of throughput of LP and HP data under the constraints of energy availability and maximum packet loss probability. Finally, we also outline some open research directions in the RF-EHNs.


\section{Overview of RF Energy Harvesting Networks}
\label{sec:overview}

In this section, we first describe a general architecture of an RF-EHN and the circuit design of an RF energy harvester. Then, we introduce the RF energy harvesting technique.

\subsection{Architecture of RF Energy Harvesting Network}

\begin{figure}
\centering
\includegraphics[width=0.95\textwidth]{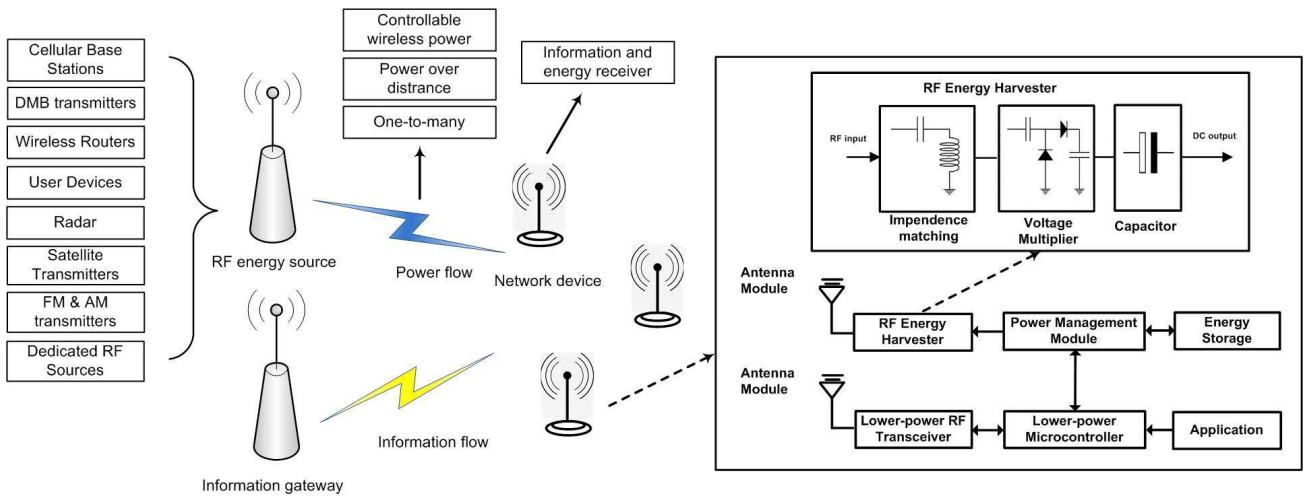}
\caption{General architecture of an RF energy harvesting network.} \label{network_architecture}
\end{figure}

A typical centralized architecture of an RF-EHN, as shown in Fig.~\ref{network_architecture}, has three major components, i.e., information gateways, the RF energy sources and the network nodes/devices. The information gateways are generally known as base stations, wireless routers and relays. The RF energy sources can be either dedicated RF energy transmitters or ambient RF sources (e.g., a TV tower). The network nodes are the user equipments that communicate with information gateways. Typically, the information gateways and RF energy sources have continuous and fixed electric supply, while the network nodes harvest energy from RF sources to support their operations. In some cases, the information gateway and RF energy source can be the same. Note that the decentralized RF-EHN also has the similar architecture as shown in Fig.~\ref{network_architecture} except that the network nodes can communicate among each other.

Figure~\ref{network_architecture} also shows the block diagram of a network node with RF energy harvesting capability. An RF energy harvesting node consists of the following major components:
\begin{itemize}
\item The application, to perform some network functions;
\item A low-power microcontroller, to process data from the application; 
\item A low-power RF transceiver, for information transmission or reception; 
\item An energy harvester, composed of an RF antenna module, an impedance matching, a voltage multiplier and a capacitor, to collect RF signals and convert them into electricity; 
\item A power management module, which decides whether to store the electricity obtained from the RF energy harvester or to use it for information transmission immediately; and 
\item An energy storage battery, to reserve the harvested RF energy for future use. 
\end{itemize}


A typical RF energy harvester consists of an antenna module, impedance matching, voltage multiplier and capacitor. Figure~\ref{network_architecture} also illustrates the block diagram of the RF energy harvester. 
\begin{itemize}
	\item The antenna module can be designed to work on either single frequency or multiple frequency bands, in which the network node can harvest from a single or multiple sources simultaneously, respectively. Nevertheless, the RF energy harvester typically operates over a range of frequencies since energy density of RF signals is diverse in frequency. 
	\item The impedance matching is a resonator circuit operating at the designed frequency to maximize the power transfer between the antenna module and the multiplier. The efficiency of the impedance matching is high at the designed frequency. 
	\item The main component of the voltage multiplier is a diode of the rectifying circuit which converts RF signals into DC voltage. Generally, higher conversion efficiency can be achieved by diodes with lower built-in voltage. The capacitor ensures to deliver power smoothly to the load. In addition, when RF energy is instantaneously unavailable, the capacitor can also serve as a reserve for a short duration. 
\end{itemize}

For the general node architecture introduced above, the network node has the separated RF energy harvester and RF transceiver. Therefore, the node can perform energy harvesting and information transmission simultaneously. In other words, this architecture supports both \emph{in-band} and \emph{out-of-band} RF energy harvesting. In the in-band RF energy harvesting, the network node can harvest RF energy from the same frequency band as information transmission. On the other hand, in the out-of-hand RF energy harvesting, the network node harvests RF energy from the different frequency band from that used for information transmission. Since RF signals can carry energy as well as information, theoretically RF energy harvesting and information reception can be performed from the same RF signal input. This is referred to as simultaneous wireless information and power transfer (SWIPT)~\cite{Varshney2008} concept. This concept allows the information receiver and RF energy harvester to operate on the same antenna module. 


Because an existing circuit is not capable of directly extracting energy from the same RF signals for information decoding~\cite{ZhangRuiMIMO}, the authors in \cite{ZhangRuiMIMO} introduce practical implementations of a co-located information receiver and energy harvester. Specifically, two receiver architectures, namely, \emph{time switching} and \emph{power splitting}, are presented. The \emph{time switching} architecture allows the network node to switch and use either the information receiver or the RF energy harvester for the received RF signals at a time. On the other hand, in the \emph{power splitting} architecture, the received RF signals are split into two streams for the information receiver and RF energy harvester with different power levels. It has been recognized that in theory \emph{power splitting} achieves better information rate and amount of RF energy harvested than those of {\em time splitting}~\cite{ZhangRuiMIMO}. However, in practice, an implementation of {\em power splitting} has higher hardware complexity than that of the {\em time splitting}. Note that the \emph{power splitting} architecture allows only in-band RF energy harvesting, while the \emph{time switching} architecture can additionally support out-of-band RF energy harvesting. 

\subsection{RF Energy Harvesting Technique}

In RF energy harvesting, the amount of harvested energy depends on the transmit power, wavelength of the RF signals and the distance between an RF energy source and the harvesting node. The amount of harvested RF energy can be calculated based on the Friis equation~\cite{Visser2013}. Unlike energy harvesting from other sources, such as solar, wind, geothermal and vibrations, RF energy harvesting has the following characteristics:

\begin{itemize}

\item RF sources can provide controllable and constant energy transfer over distance for RF energy harvesters, especially for a fixed RF-EHN.


\item RF energy harvesting technique is suitable for mobile (handheld) devices.

\item Since the amount of harvested RF energy depends on the distance from the RF source, the network nodes in the different locations can have significant difference in the amount of harvested RF energy. 

\end{itemize}
 
With RF energy harvesting and transfer, proactive replenishment at mobile devices, rather than passive adoption to the environmental resources, can be achieved which is more suitable for applications with QoS requirements. 

The RF sources can mainly be classified into two types: dedicated RF sources and ambient RF sources. Dedicated RF sources can be deployed to provide energy to network nodes when more predictable energy supply is needed. Ambient RF sources are the RF transmitters that are not intended for RF energy transfer (e.g., TV and radio tower). This RF energy is essentially free. Ambient RF sources can be static or dynamic. The study in~\cite{SLee2013} is an example of energy harvesting from dynamic ambient RF sources in a cognitive radio network. A secondary user can harvest RF energy from nearby transmitting primary users, and it can transmit data when it is sufficiently far from primary users or when the nearby primary users are idle.

It is reported in \cite{Visser2013} that, in between $25$ and $100$ meters from a base station, the aggregated power density over the GSM900 downlink frequency band ranges from $3.0$ to $0.1$ $mW/m^{2}$ indoors everywhere or outdoors on a high level. While the power density received from GSM1800 downlink is in the same order of magnitude as those received from GSM900 frequency band. Also, a state-of-the-art prototype implementation in \cite{V2013Liu} is shown to achieve an information rate of $1$ kbps between two wireless devices powered by ambient RF signals, at distances of up to $2.5$ feet and $1.5$ feet for outdoors and indoors, respectively. Tested at a variety of locations, the implemented end-to-end system is able to operate battery-free at distances of up to $6.5$ miles from the TV tower.

\subsection{Existing Applications of RF Energy Harvesting}

Wireless sensor networks have become the most widely used applications of RF-EHNs. An RF energy harvester can be used in a sensor node to supply energy. For example, \cite{Nishimoto2010} presents a prototype implementation of sensor nodes powered by ambient RF energy. The RF-EHN also has attractive healthcare and medical applications such as wireless body network. Benefiting from RF energy harvesting, low-power medical body sensors can achieve real-time work-on-demand power, which further enables a battery-free circuit, reducing the size of the nodes.
In~\cite{Zhang2010}, the authors design the RF-powered energy-efficient application-specific integrated circuit, fabricated in standard $0.18\text{-}m$ CMOS technology and featured with a work-on-demand protocol. The integrated circuit is for wireless body networks in medical applications. 
Furthermore, RF energy harvesting can be used to provide charging capability for a wide variety of low-power mobile devices such as electronic watches, hearing aids, and MP3 players, wireless keyboard and mouse, as most of them consume only micro-watts to milli-watts range of power. 

\section{Design Issues of Resource Allocation in Radio Frequency Energy Harvesting Networks}
\label{sec:issues}

RF-EHNs introduce RF energy harvesting as a new function for wireless devices. As a consequence, resource allocation in RF-EHNs has to take into account two different objectives, i.e., information transmission/reception and RF energy harvesting.
These issues are receiver operation policy, beamforming, medium access control (MAC) protocol, cooperative relaying and routing protocol. Moreover, we review the state-of-art design approaches attempted to address these issues.

\subsection{Receiver Operation Policy}



A receiver operation policy is required for wireless nodes sharing the same antenna or antenna array for information reception and RF energy harvesting. The policy can be designed to deal with various tradeoffs in the physical layer and MAC layer to meet certain performance goals. Most of the existing policies are either based on \emph{time switching} or \emph{power splitting} architecture. The focus of the \emph{time switching} architecture is to coordinate the time for information reception and RF energy harvesting. On the other hand, for the \emph{power splitting} architecture, the operation policy is to find an optimal ratio to split the received RF signals.

The authors in \cite{IKrikidis2012} study a simple greedy switching policy based on the \emph{time switching} architecture. The idea of the policy is to let the relay node transmit when its remained energy can support information transmission. Through simulation, the greedy switching policy is shown to achieve a near optimal performance over a wide range of signal-to-noise ratio (SNR). In \cite{SDurrani2013}, the authors consider a three-node amplify-and-forward network with an RF energy harvesting relay. Two relaying protocols for the relay node are proposed based on the \emph{time switching} and \emph{power splitting} architectures. Specifically, the authors derive the optimal RF energy harvesting time for the \emph{time switching} based relaying protocol and the optimal value of power splitting ratio for the \emph{power splitting} based relaying protocol. The evaluation results indicate that the \emph{time switching} based relaying protocol is superior in terms of throughput at relatively low SNR and high transmission rates. However, as the transmit power considered is variable, it incurs considerable hardware complexity.
\cite{X2014Lu} studies receiver operation policy in a multiple-channel cognitive radio network, in which the secondary user select channels not only for information transmission but also for energy harvesting. In the context of complete CSI at the secondary user, the optimal policy for the secondary user to maximize throughput is determined, based on the remaining energy level and the number of waiting packets in data queue, by applying an MDP-based optimization.



\subsection{Beamforming}
A key concern for RF information and energy transfer is the decay in energy transfer efficiency with the increase of transmission distance due to propagation path loss. Multi-antenna techniques can be used to achieve spatial multiplexing. Furthermore, beamforming techniques employing multiple antennas can be applied to achieve improved efficiency of RF energy transfer \cite{X2013Chen} as well as SWIPT \cite{ZhangRuiMIMO}, without additional bandwidth or increased transmit power. 
A problem arising in beamforming is channel state estimation feedback. Designing a feedback mechanism is challenging in RF-EHNs because existing channel training and feedback mechanisms used for an information receiver are not applicable for an energy harvester due to the hardware limitation.
 


Beamforming is first explored in a three-node multiple-input multiple-output (MIMO) network~\cite{ZhangRuiMIMO} with one transmitter, one energy harvester and one information receiver. The authors in \cite{ZhangRuiMIMO} study the optimal transmission strategies to achieve tradeoff between information rate and amount of RF energy transferred under the assumption of perfect knowledge of channel state information at the multi-antenna transmitter employing beamforming. 
More recently, \cite{X2013Chen} considers to utilize energy beamforming in a large-scale MIMO system to improve energy efficiency in long-distance power transfer. A resource allocation scheme is proposed to jointly optimize the transmit power and duration of RF energy transfer for maximizing energy efficiency under some QoS requirement.  
Beamforming has also been advocated to provide secure communication in SWIPT systems with eavesdroppers. For example, in \cite{Ng1311.2507}, the authors aim to protect transmitted information to the intended receiver by jointly generating artificial noise to the eavesdroppers through beamforming. A non-convex optimization problem for beamforming design is formulated to minimize the total transmit power, under the requirements for both information transmission and artificial noise generation.


\subsection{MAC Protocol}

To achieve QoS support and fairness for information transmissions, a MAC protocol specifically designed for RF-EHNs is needed to coordinate the network nodes' transmissions. In addition to the channel access for information transmission, the network nodes also need to spend some time for RF energy harvesting. The challenge is that the time taken to harvest enough energy is different for different nodes due to various factors such as types of the available RF energy sources and distance. The MAC protocols coordinate network nodes either in a contention-free approach (e.g., polling) or a contention-based approach (e.g., CSMA/CA). The contention-free MAC protocol needs to take the node-specific RF energy harvesting process into account to achieve high throughput and fairness. With the contention-based MAC protocol, each node contends for radio resources for information transmission. If the RF energy harvesting duration is not optimally decided, an extended delay of resource contention due to communication outage may incur.

In \cite{Kim2011}, the authors present an energy adaptive MAC protocol for RF-EHNs. Two energy adaptive methods, i.e., energy adaptive duty cycle and energy adaptive contention algorithms, are proposed to use the node energy harvesting status (e.g., RF energy harvesting rate) as a control variable to manage the node's duty cycle and backoff time, respectively. However, the energy adaptive MAC protocol requires centralized control and out-of-band RF energy supply. In contrast, the authors of \cite{Nintanavongsa2013} consider in-band RF energy supply. A CSMA/CA-based MAC protocol called RF-MAC is designed to optimize RF energy delivery rate to meet the energy requirement of sensor nodes while minimizing disruption to data communication. The RF-MAC also incorporates the methods to select RF energy sources. Also, the energy and data rate tradeoff is analyzed.

\subsection{Cooperative Relaying}

Cooperative relaying can improve the network performance in terms of efficiency and reliability by using intermediate relay nodes. Hence, it is particularly suitable to be applied in energy constrained networks like RF-EHNs. 
Relay selection is a decision factor in the performance of cooperative relaying. The main challenge lies in that the preferable relay for information transmission does not necessarily coincide with the relay has the strongest channel for energy harvesting. Thus, as a tradeoff, relay selection has to leverage between the efficiency of information and
energy transfer. To this end, other than taking channel state information into account, information about energy status (e.g., internal energy reserve and potential external RF energy arrival) must also be regarded, which makes relay selection more complex. 
 
In \cite{Michalopoulos2013}, the authors study two relay selection schemes, i.e., the time-sharing selection and the threshold-checking selection. In the time-sharing selection, the source node switches among the relays with the maximum SNR. In the threshold-checking selection, the source node chooses the relay with the highest RF energy harvesting rate. It is demonstrated that the threshold-checking selection has better performance in terms of achieved capacity for the given RF energy harvesting requirement. On the other hand, the time-sharing selection has better performance in terms of outage probability when the normalized average SNR per link is larger than $5$dB. \cite{I2014Krikidis} investigates relay selection from a system perspective. Specifically, the authors examine a random relay selection policy based on a sectorized area with central angle at the direction of each receiver. A geometry approach is adopted to study the impact of cooperative density and relay selection in a large-scale network with SWIPT.


\subsection{Routing Protocol}

An RF-EHN can be based on multihop transmission, where routing is a crucial issue. Unlike the energy-aware routing developed in conventional wireless networks, the routing protocols in the RF-EHN must take the RF energy propagation and the circuit design of network nodes (e.g., RF energy harvester's sensitivity) into account. This is due to the fact that the amount of harvested RF energy available at each node can be different. In addition, the routing metric has to be jointly defined based on RF energy harvesting parameters (e.g., RF signal density, energy conversion rate, and distance from RF sources) and network parameters (e.g., link quality and number of hops).

In \cite{Doost2010}, the authors consider the routing problem in a wireless sensor network where the sensor nodes are charged wirelessly with in-band RF energy. It is shown that simple metrics such as a hop count may not be suitable for routing in such networks. Therefore, a new routing metric based on the charging time of the sensor nodes is introduced. Then, the modified Ad hoc On-Demand Distance Vector (AODV) routing protocol considering the new routing metric is proposed. In this protocol, the sensors choose the path with the lowest value of maximum charging time. %

\section{Optimization Design for Mobile Energy Harvesting Node with Delay-limited Communication: A Case Study}
\label{sec:optimization}

In this section, we show a case study of a receiver operation problem with joint QoS support and service differentiation in an RF-EHN, which, to the best of our knowledge, has not been addressed before.


\subsection{System Model}

We consider a node having two types of data, i.e., LP and HP data. The arriving packets (e.g., from different applications at higher layers) are stored in two separate queues. Both of the queues have finite sizes. There is an access point (AP) which performs as a information gateway and dedicated RF energy source. The node in a coverage area of the AP can decide to request for RF energy transfer from the AP or to transmit a packet of LP or HP data to the AP. Specifically, the node adopts the \emph{time switching} receiver architecture as proposed in~\cite{ZhangRuiMIMO}. That is, the node works either in an RF energy harvesting mode or an information transmission mode. Besides, a battery, which has a finite capacity, is equipped with the node to store energy harvested from the AP. We consider the packet loss requirement for each type of data. In this case, a packet is dropped (i.e., loss) if the corresponding queue is full or the battery is empty upon its arrival. The LP and HP data could have different maximum packet loss probability requirements. 
   
\subsection{Optimization Problem}
   
When the node is in the coverage area of an AP, the node is facing a decision making problem (i.e., receiver operation) whether to request and harvest RF energy or to transmit a packet from the queue of LP or HP data to the AP, based on the \emph{time switching} architecture. The decision making problem must be solved to achieve the objective (i.e., maximizing weighted sum of throughput of LP and HP data) and meet the QoS requirements taking the following factors into account. 
\begin{itemize}
	\item \emph{Packet arrival}: The node has independent packet arrivals for HP and LP data. The probabilities of $a$ packets arriving at the node for the LP and HP data are denoted as $\alpha_a$ and $\lambda_a$, respectively.
	\item \emph{Packet transmission}: If the node decides to transmit a packet retrieved from the queue of either LP or HP data to the AP, the successful packet transmission probability is denoted as $\mu$. The transmission of one data packet by the node consumes $K$ units of energy from the battery. If the battery has energy less than $K$ units, the node cannot transmit the packet.
	\item \emph{RF energy harvesting and transfer}: If the node decides to request for RF energy from the AP, the node can harvest $w$ units of energy (i.e., the energy level of the battery increases by $w$ units) successfully with the probability $\sigma_w$. Note that this parameter can be adopted from experiments.
\end{itemize}

\subsection{Optimization Formulation}

To achieve the objective in terms of the throughput and to meet the QoS requirement in terms of packet loss probability, we formulate an optimization model to obtain the optimal operation policy for the node based on constrained Markov decision process. The optimal operation policy determines the action to be taken by the node given its current state.

\subsubsection{State Space and Action Space}

The state space of the node is defined by the possible energy level in the battery, and the numbers of packets in the queues for LP and HP data. For the node, the possible action (i.e., the action space) will be to transmit a packet from the queue of LP or HP data to the AP, or to request for RF energy from the AP.

The state transition happens in the following events:
\begin{itemize}
	\item {\em Data Transmission:} The energy level of the battery will reduce by $K$ units and the number of packets in the queue of HP or LP data decreases by one (i.e., depending on which type of data that the node decides to transmit) with probability $\mu$. These state transitions happen given that the queue of the HP or LP data selected by the node is not empty.
	\item {\em Request for RF energy transfer:} The energy level of the battery increases by $w$ units with probability $\sigma_w$.
	\item {\em Packet arrival:} The numbers of packets in the queues of HP and LP data increase by $a$ with probabilities $\alpha_a$ and $\lambda_a$, respectively.
\end{itemize}

\subsubsection{Optimal Operation Policy}

The mapping of a state to an action taken by the node is referred to as the policy denoted by $\pi$. The optimal operation policy is defined to achieve the maximum long-term average weighted sum of throughput of the LP and HP data, while the packet loss requirements of LP and HP data are maintained below the thresholds. The objective function of the optimization model is expressed as follows:
\begin{align}
	\max_{\pi}: & \quad \mathscr{J}_{T}(\pi) = \lim_{t \to \infty} \inf\frac{1}{t} \sum^{t}_{t^{\prime}=1} \mathbb{E}( \omega_{\mathrm{LP}}\widetilde{\mu}_{l,t^{\prime}}+\omega_{\mathrm{HP}}\widetilde{\mu}_{h,t^{\prime}} ) 
\label{primary} 
\end{align}
where $\mathscr{J}_{T}(\pi)$ is the function of weighted sum of throughput, $\omega_{\mathrm{LP}}$ and $\omega_{\mathrm{HP}}$ are the weights of LP and HP data, respectively. $\widetilde{\mu}_{ \mathrm{LP} ,t^{\prime}}$ and $\widetilde{\mu}_{ \mathrm{HP} ,t^{\prime}}$ are the successful packet transmission probabilities for LP and HP data at time $t^{\prime}$, respectively. We have $\widetilde{\mu}_{ \mathrm{LP} ,t^{\prime}}=\mu$, if the node is transmitting a packet from the queue of LP data (i.e., the queue of LP data is not empty) and there is sufficient energy in the battery for transmitting the packet (i.e., the energy level of the battery is greater than or equal to $K$). Otherwise, we have $\widetilde{\mu}_{ \mathrm{LP} ,t^{\prime}}=0$. Similarly, we have $\widetilde{\mu}_{ \mathrm{HP}, t^{\prime}}=\mu$, if the node is transmitting a packet from the queue of HP data (i.e., the queue of HP data is not empty) and there is sufficient energy in the battery. Otherwise, we have $\widetilde{\mu}_{ \mathrm{HP}, t^{\prime}}=0$. It is worth noting that the data transmission is constrained by RF energy harvesting, i.e., data transmission is successful provided that there is sufficient harvested energy in the battery. 

Let us consider LP data. The constraint of a packet loss probability requirement is expressed as follows:
\begin{equation}
	{\mathscr{J}}_{\mathrm{LP}}(\pi)	=	\lim_{t \rightarrow \infty} \sup \frac{1}{t} \sum_{t'=1}^{t} {\mathbb{E}} \left( {\mathscr{L}}_{\mathrm{LP}} \right)	\leq	L_{\mathrm{LP}}
\end{equation}
where $L_{\mathrm{LP}}$ is the packet loss requirement for the LP data, and ${\mathscr{L}}_{\mathrm{LP}}$ is the immediate packet loss probability. The immediate packet loss probability is $\mathscr{L}_{\mathrm{LP}}=\frac{\sum^{A}_{a=Q_{\mathrm{LP}}-q_{\mathrm{LP}}+1} \alpha_a}{ \bar{\alpha} }$, if there is not enough space in the queue for LP data, whose maximum capacity is denoted by $Q_{\mathrm{LP}}$. $q_{\mathrm{LP}}$ is the current number of packets in the queue. $A$ is the maximum number of arriving packets. $\bar{\alpha}$ is the average packet arrival rate of LP data. The immediate packet loss probability for the HP data can be obtained in a similar way.

Due to the space limit, we omit the derivation of the transition probability matrix. The detailed derivation of the Markov decision process based optimization problem is similar to the approach in \cite{X2014Lu}. 
To obtain the optimal operation policy of the node, we can apply a standard method to solve the constrained Markov decision process~\cite{puterman_1994}.

\subsection{Performance Evaluation}

\subsubsection{Parameter Setting}
 
The node has the battery with the size of $50$ units of energy. The maximum queue sizes for LP and HP data are $4$ packets. Unless otherwise stated, the packet arrival probabilities for LP and HP are 0.15. The successful packet transmission probability of the node to the AP is $0.99$. The probability of successful RF energy transfer and harvesting is $0.98$. If the RF energy harvesting is successful, the node will receive $4$ units of energy. There is no packet loss probability requirement for the LP data, but it is $0.1$ for the HP data. For the comparison purpose, we consider a static policy in which the node chooses three actions with equal probabilities. 
 
\subsubsection{Numerical Result}
 
\begin{figure*}[t]
\begin{center}
$\begin{array}{ccc} 
\epsfxsize=2.2 in \epsffile{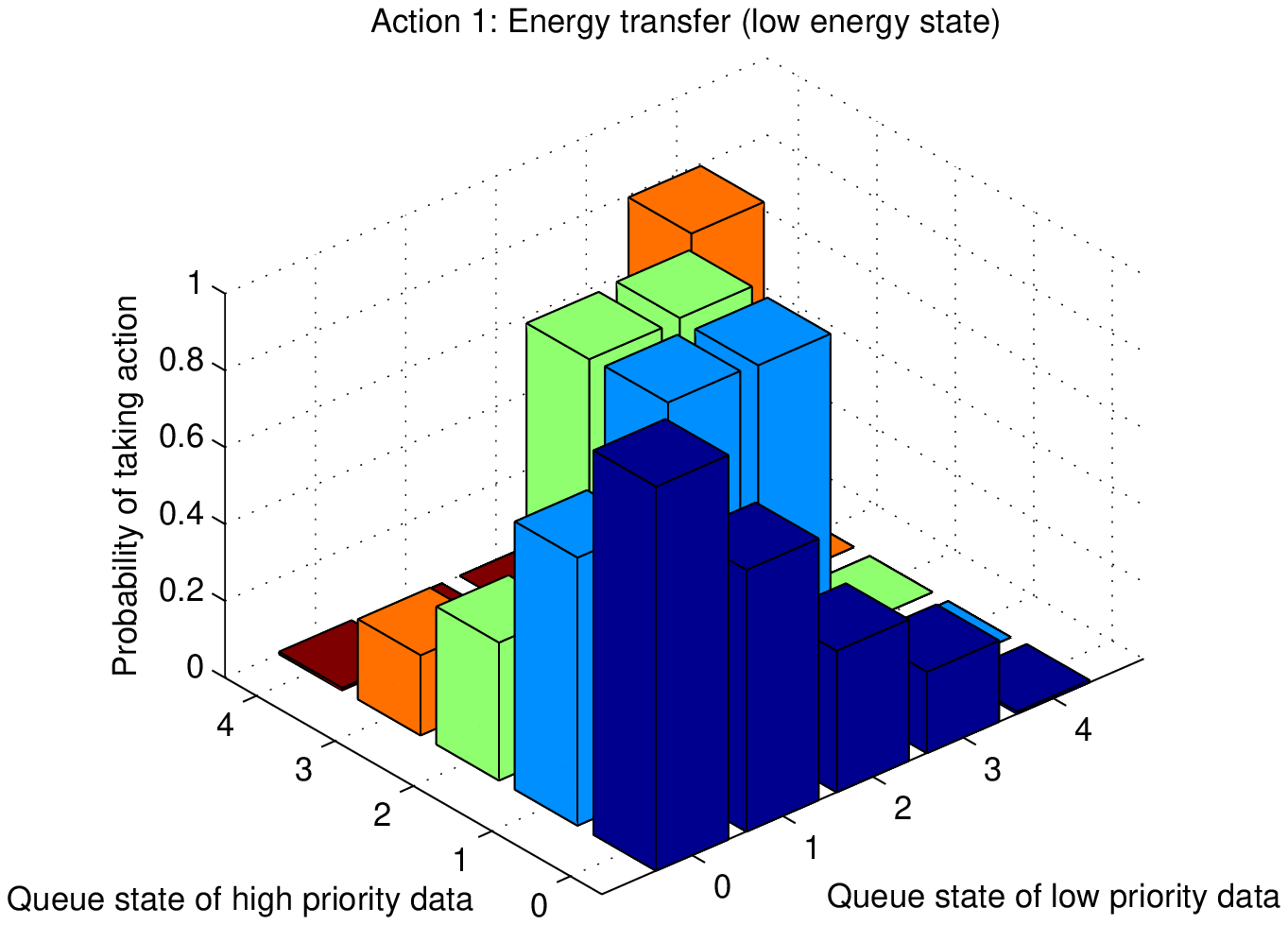}	&
\epsfxsize=2.2 in \epsffile{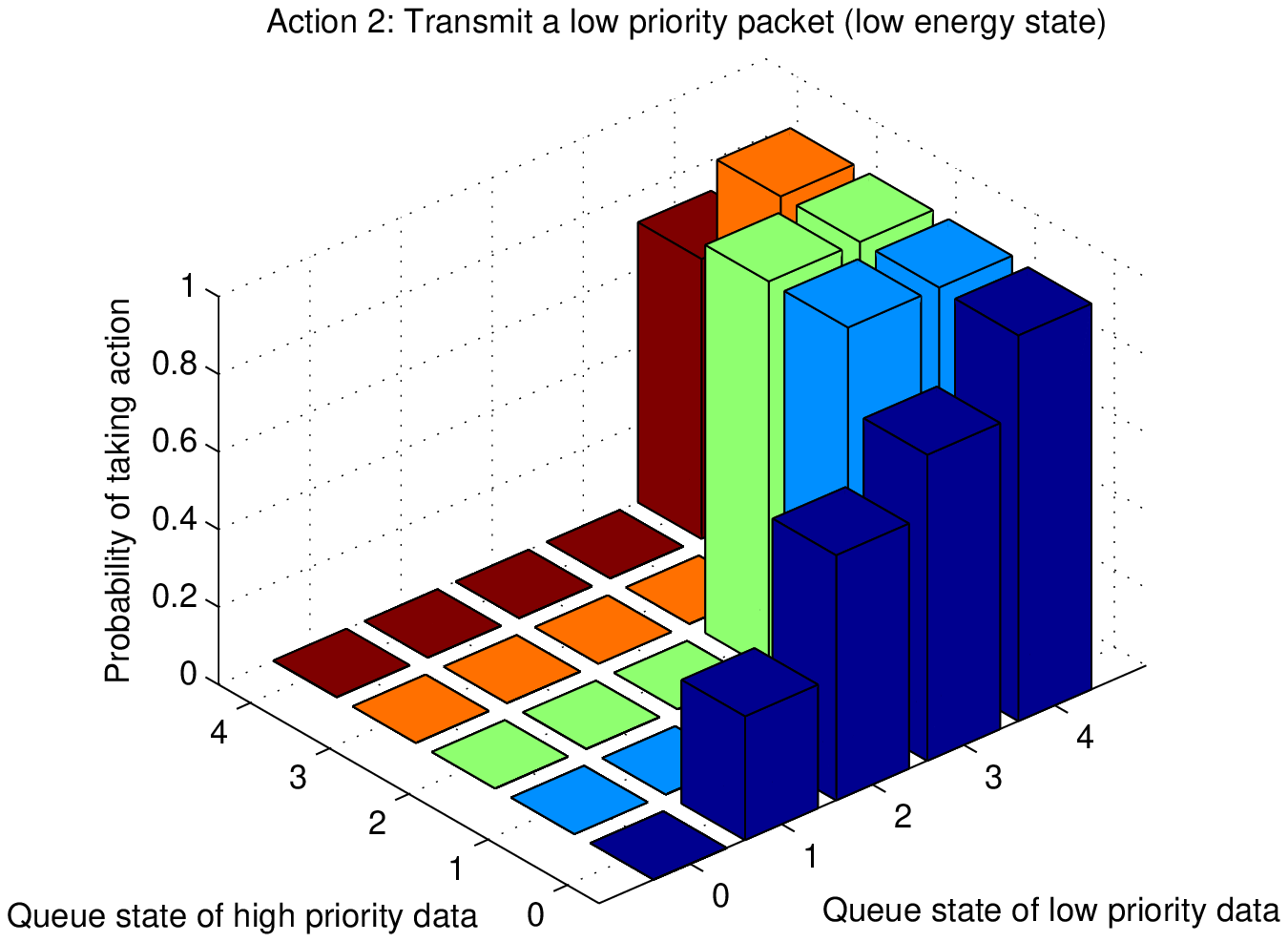}	&
\epsfxsize=2.2 in \epsffile{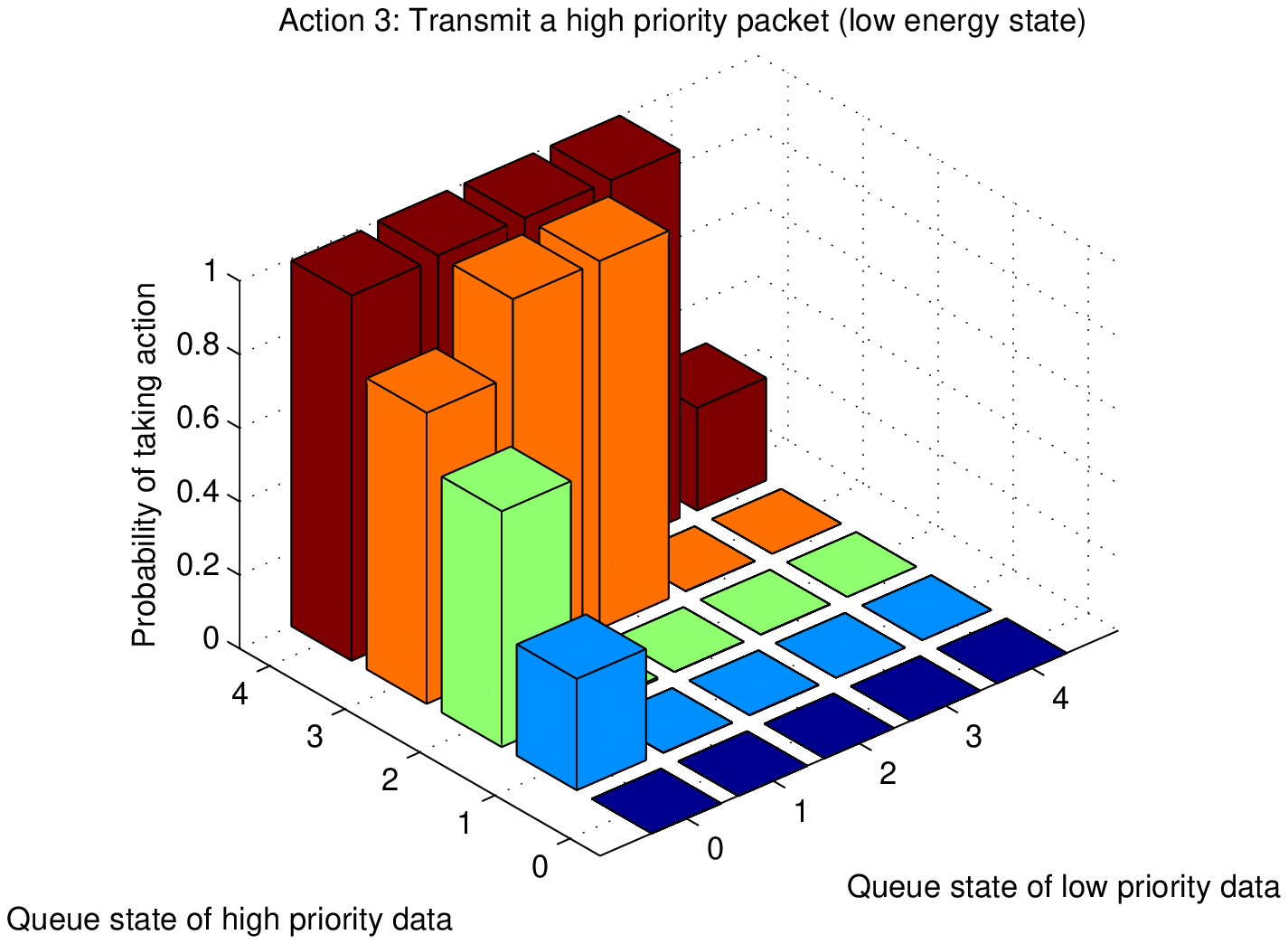}	\\ [-0.2cm]
(a)	& (b)	&	(c)
\end{array}$
\caption{Optimal operation policy for (a) requesting for RF energy, (b) transmitting a packet from the queue of LP data, and (c) transmitting a packet from the queue of HP data.}
\label{Figure5}
\end{center}
\end{figure*}

We first examine the optimal operation policy of the node obtained from solving the optimization model, when the energy level of the battery of the node is low (i.e., 5 units). The results shown in Fig.~\ref{Figure5} are obtained by solving the optimization problem (\ref{primary}) under the constraints of packet lost probability requirement for both LP and HP data.  Figure~\ref{Figure5}(a) shows that when the number of packets in the queue is small, the node tends to request for RF energy. On the other hand, when the number of packets in both queues is high, the node also tends to request for RF energy because of the high demand of energy for data transmission. As the numbers of packets waiting in the queues grow, the node will be likely to transmit a packet especially for HP data to avoid violating the packet loss requirement. Figures~\ref{Figure5}(b) and (c) show such a policy for packet transmission. 


%

\begin{figure}
\centering
\includegraphics[width=0.5\textwidth]{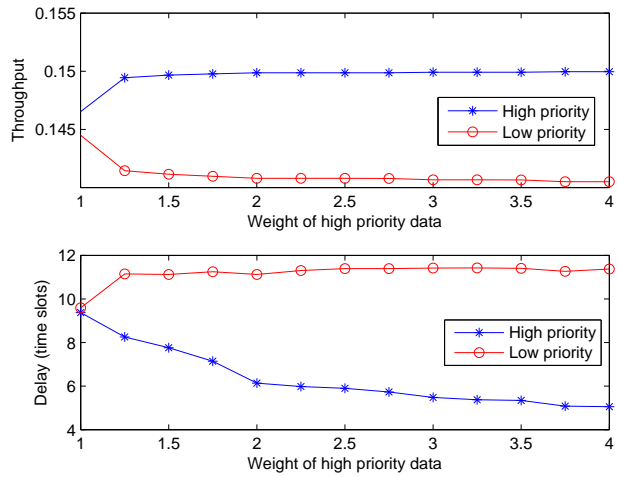}
\caption{Throughput and delay under different weight of high priority data} \label{weight}
\end{figure}

In Fig.~\ref{weight}, we vary the weight of the HP data and show the corresponding throughput and packet delay. We observe that when the weight of HP data is large, the optimal operation policy yields more opportunity for the node to transmit a packet from the queue of HP data. As a result, the throughput of the HP data increases, while that of the LP data decreases. Correspondingly, when the weight of HP data increases, the delay of the LP data increases, while that of the HP data decreases. To achieve the performance goal, the weights of the optimization model can be adjusted. It is interesting to observe the unbalanced performance improvement and degradation of the HP and LP data, respectively, when the weight is adjusted. Specifically, while the throughput and delay of the HP data improve slightly, those of the LP data degrade significantly. This is due to the fact that the node has to reserve energy resource by not transmitting a packet from the queue of LP data too much so that the energy can be used for the packet transmission of the HP data in the future.

\begin{figure}[h]
\begin{center}
$\begin{array}{c} \epsfxsize=3.1 in \epsffile{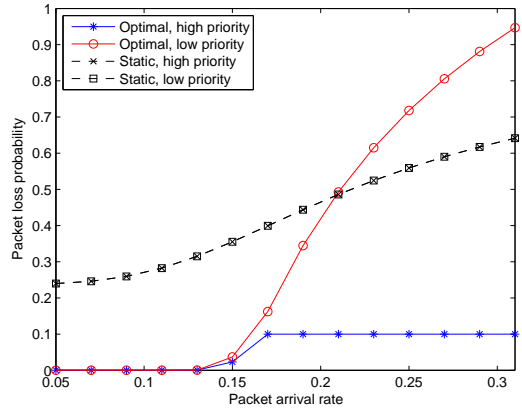} \\ [-0.2cm]
\end{array}$
\caption{Packet loss probability under different packet arrival rates.} 
\label{fig:vary_arr}
\end{center}
\end{figure}

Furthermore, we study the packet loss behavior of the proposed operation policy. For the comparison purpose, we consider a static policy in which the node chooses three actions with equal probabilities. 
Figure~\ref{fig:vary_arr} shows the packet loss probability when the packet arrival rates of both the HP and LP data are varied. As expected, when the packet arrival probabilities increase, the packet loss probabilities increase. However, at a certain point, the optimal operation policy successfully maintains the packet loss probability of the HP data at the defined threshold, which is 0.1, while that of LP data increases and becomes unbounded. In addition, we also show the results from the static policy, which fails to achieve acceptable performance, specially for the HP data.

The studied model can be extended to the case of multiple nodes where a scheduling policy is required to coordinate the uplink data transmission and downlink energy transfer. The operation policy of nodes has to take the scheduling policy into account. 

\section{Open Research Issues}
\label{sec:openissues}

In this section, we discuss about some open research issues as followings.
\subsection{Technological Directions}
 
\begin{itemize} 

\item {\em Distributed Energy Beamforming}:
Distributed energy beamforming enables a cluster of distributed energy sources to cooperatively emulate an antenna array by transmitting RF energy simultaneously in the same direction to an intended energy receiver for better diversity gains. The potential energy gains at the receiver from distributed energy beamforming are expected to be the same as that from the well-known information beamforming. However, challenges arise in the implementation, e.g., time synchronization among energy sources and coordination of distributed carriers in phase and frequency so that RF signals can be combined constructively at the receiver. 

\item {\em Cooperative Sensing and Spectrum Sharing}: 
In contrast to a conventional cognitive radio network, spectrum occupancy by primary users is not necessarily undesirable in RF-EHNs, as it results in RF energy harvesting opportunities. The cooperative spectrum sensing and sharing techniques for cognitive radio networks can be directly adopted to help secondary users with RF energy harvesting capability to identify the occupied spectrum bands and RF energy harvesting opportunity. However, because of the dispersed geographic locations of the secondary users, they may experience different spectrum conditions due to different activities and locations of primary users. High utilization of frequency band and high efficiency in detecting frequency usage require information exchange and fusion among secondary users, which can be a challenging task in network design.  
  
\item {\em Interference Management}:
Existing interference management techniques, e.g., interference alignment and interference cancellation, attempt to avoid or mitigate interference through spectrum scheduling. However, with RF energy harvesting, harmful interference can be turned to useful energy through a scheduling policy. In addition, the scheduling policy can be combined or integrated with power management schemes for further improvement in energy efficiency.

\item {\em Energy Trading}:
In RF-EHNs, RF energy becomes a valuable resource. The RF energy market can be established to economically manage this energy resource jointly with radio resource. For example, wireless charging service providers may act as RF energy suppliers to meet the energy demand from network nodes. The wireless energy service providers can decide on pricing and guarantee the quality of charging service. The key feature of this market is on-demand trading where the wireless charging service providers offer charging service to the network nodes on a real-time and on-demand basis. One of the efficient approaches in this dynamic market is to develop demand side management, which allows the service providers and network nodes to interact like in smart grid, to guarantee energy-efficiency and reliability. However, the issues related to the amount of RF energy and price at which they are willing to trade while optimizing the tradeoff between the revenue and cost must be investigated. 
\end{itemize}

\subsection{Application Directions} 

\begin{itemize}
\item {\em Wireless Machine-to-machine (M2M) Communications}
Deploying massive and unmanned wireless M2M devices introduces the new challenge: how could such a huge amount of devices be powered? RF energy harvesting provides an alternative solution. For example, the ``last meter'' technologies, e.g., WiFi, IEEE 802.15, ZigBee and UWB, can be potentially used for RF energy supply for wireless M2M devices.


\item {\em Vehicular Communications}
Vehicular transmitters in vehicular networks (i.e., vehicle-to-vehicle and vehicle-to-infrastructure communications) can be energy sources for wireless devices belonged to passengers or even pedestrians.
For example, the passengers' devices can harvest RF energy from an on-board unit deployed in a bus. Alternatively, when the passengers are at the bus stop, their devices can also harvest RF energy from roadside units.

\item {\em Smart Automation}:
In automation systems, RF energy harvesting can eliminate the wired connection for power supply for sensor and actuator devices, especially for the devices installed on the moving components, e.g., wheel and rotating assemblies. For example, the turbine blade sensors on mechanical engines can be steadily provisioned by dedicated RF sources to maintain the function of monitoring on the blades' status.

 
\item {\em Device-to-device (D2D) Communications}:  D2D communications, underlaid or overlaid with cellular networks, allows user equipment to access the same spectrum band for cellular communication with interference constraints. The occupied spectrum provides RF energy harvesting opportunities, especially when network density is high. The user equipment can harvest and use RF energy for their local direct D2D communications.
\end{itemize} 

\section{Conclusion}
\label{sec:conclusion}
 
Radio frequency (RF) energy harvesting and transfer techniques play an important role in powering the next generation of wireless networks. In this article, we have presented an overview of the RF energy harvesting networks (RF-EHNs), including the network architecture and the enabling techniques. We have introduced the major design issues in resource allocation of the RF-EHNs, and reviewed some up-to-date research progresses. Moreover, we have shown a case study on how to design a QoS-aware receiver operation policy with service differentiation in a general RF-EHN. We obtain an optimal operation policy to maximize the throughput of a mobile node with RF energy harvesting capability, as well as to provide service differentiation among two different types of data, under the constraints of packet loss probability. In addition, we foresee the future research directions of the RF-EHNs.

\end{document}